\documentstyle[aps,prl,twocolumn,epsf]{revtex}

\begin{document}
\draft
\twocolumn[\hsize\textwidth\columnwidth\hsize\csname@twocolumnfalse\endcsname
\preprint{}
\title{ Experimental evidences of Luttinger liquid behavior
in the crossed multi-wall carbon nanotubes}
\author{ Jinhee Kim$^1$, Kicheon Kang$^2$, Jeong-O Lee$^2$,
       Kyung-Hwa Yoo$^1$, Jae-Ryoung Kim$^2$, Jong Wan Park$^1$,
       Hye Mi So$^2$, and Ju-Jin Kim$^2$ }
\address{$^1$Electricity Group, Korea Research Institute of Standards
         and Science, Yusong-gu P.O.Box 102, Taejon, 305-600, Korea }
\address{$^2$Department of Physics, Chonbuk National University,
         Chonju 561-756, Korea}
\date{\today}
\maketitle
\begin{abstract}
  Luttinger liquid behavior was observed in a crossed junction formed with
  two metallic multi-wall carbon nanotubes
whose differential conductance vanished with the power of bias voltage
and temperature. With applying constant
voltage or current to one of the two carbon nanotubes in a crossed geometry,
the electrical transport properties of
the other carbon nanotube were affected significantly, implying there
exists strong correlation between the carbon
nanotubes. Such characteristic features are in good agreement with
the theoretical predictions for the crossed two
Luttinger liquids.

\end{abstract}
]
\narrowtext

The non-Fermi liquid behavior of low dimensional systems is one of
the most challenging subjects in recent theoretical and
experimental studies~\cite{tarucha,bockrath,komnik,egger}. In one
dimension, electron-electron interaction is known to invalidate
the Fermi liquid description of metal and gives a new physical
state, Luttinger liquid (LL), which is characterized by the
spin-charge separation and  the absence of single-particle
excitation at low energy, etc. Until now, many experiments, mostly
with the compound semiconductors~\cite{tarucha}, have been performed
regarding the LL behavior of low dimensional conductor. The
experimental results seem to support non-Fermi liquid behavior of
one-dimensional conductor but conclusive evidence of the LL
behavior is still to come.

Carbon nanotube (CNT)~\cite{dekker} is known to be an ideal system
to test the LL behavior of one-dimensional
system~\cite{bockrath,komnik,egger}. Previous experiment on
single-wall CNT by Bockrath {\em et al.}~\cite{bockrath} has shown
that the tunneling density of states vanishes both with
temperature and bias voltage in the power-law functional form.
However, the electrical transport properties of single-wall CNT in
their measurements were dominated by the Coulomb blockade effect
at low bias region, which might mask the LL behavior.
Recently, Komnik and Egger~\cite{komnik} have proposed an elegant
way to verify the LL behavior of one-dimensional conductor. They
have shown that if two LLs are contacted in a point-like manner,
the electrical transport through one LL is perturbed significantly
by the bias voltage applied to the other LL. A previous experiment
on the crossed single-wall CNT~\cite{fuhrer} did not show the LL
behavior theoretically predicted~\cite{komnik}, probably because
of the large Coulomb blockade effect in each CNT. To test these
theoretical predictions and detect possible deviation from the
Fermi liquid theory, we have fabricated a cross junction formed
with two metallic multi-wall CNTs. With single-wall CNTs, it is
not easy to form a cross junction with enough electrostatic interaction
between two CNTs which is essential to observe LL behavior 
in the crossed geometry~\cite{komnik}. Due to the relative 
easiness of forming low-ohmic cross junction and weak Coulomb
blockade effect compared
to the single-wall counterpart, the multi-wall CNT has great
advantages to test the LL behavior of CNT. Further, it is
important to note that multi-wall CNT, as well as single-wall CNT,
is predicted to show LL behavior in the limit of small number of
conducting channels~\cite{egger}.

In this Letter, we report experimental evidences of LL behavior of
multi-wall CNTs in the crossed geometry. Each CNT in the sample
showed vanishing tunneling density of states with bias voltage
in the power-law functional form, which is one of the signatures
of LL behavior. The differential conductance curves at different
temperatures are collapsed well into a single scaling curve. We
have also measured the differential conductance of one CNT with
applying constant voltage to the other CNT and observed that the
differential conductance of one CNT is strongly influenced by the
voltage applied to the other CNT ~\cite{komnik}. Furthermore, it
is found that the off-diagonal differential resistance curve in
the cross junction shows highly non-linear behavior along with
negative differential resistance, which is another indication of
the existence of strong correlation between two CNTs. All the
experimental results support the LL behavior of multi-wall CNTs.

The multi-wall CNT used in this measurement was synthesized by arc
discharge method. To select single CNT we have dispersed
ultrasonically the CNT in chloroform for about half an hour and
then dropped a droplet of the dispersed solution on the Si
substrate with 500 nm-thick thermally-grown $\mbox{\rm SiO}_2$
layer. The multi-wall CNTs in the crossed form were searched by
scanning electron microscope (SEM). The patterns for electrical
leads were generated using e-beam lithography technique onto the
selected CNTs and then 20 nm of Ti and 50 nm of Au were deposited
successively on the contact area by thermal evaporation. Shown in
Fig. 1 is the SEM photograph of the crossed CNTs with the electric
leads labeled. The atomic force microscope study has shown that
the diameter of the CNT was in the range of 25 - 30 nm. In order
to form low-ohmic contacts between the CNT and the Ti/Au
electrodes, we have performed rapid thermal annealing at 800 C
for 30 sec~\cite{lee}. The contact resistances are in the range of
5 k$\Omega$ - 18 k$\Omega$ at room temperature and become 10
k$\Omega$ - 60 k$\Omega$ at 4.2 K. The cross junction has junction
resistance of 5.4 k$\Omega$ at room temperature and 16.8 k$\Omega$
at 4.2 K. The four-terminal resistance of each CNT increases
monotonically with lowering temperature and depends sensitively on
the bias current level, implying non-ohmic current-voltage
characteristics of the CNT~\cite{lee}.

We have measured the current-voltage ($I$-$V$) characteristics
both in two- and four-terminal measurement configurations. The
differential conductance-voltage curve ($dI/dV$-$V$) was then
obtained by numerically differentiating the $I$-$V$
characteristics. Insets of Fig. 2 display the four-terminal
$dI/dV$-$V$ curves of each CNT as a function of temperature. Subtracting
the contact resistance, both two- and four-terminal measurements give
nearly identical $dI/dV$-$V$ curves. As shown in the insets of
Fig. 2, the differential conductance, which is proportional to the
density of states near the Fermi level, vanishes at low bias as
the temperature is lowered. We have found that the density of
states vanishes with the bias voltage in the power-law functional
form, $G\sim V^\alpha$, with $\alpha=0.3$ for the nanotube
horizontally placed in Fig. 1 (from now on we call it CNT-1) and
$\alpha=0.9$ for the nanotube vertically placed (CNT-2). By using
the relation between the exponent and the Luttinger parameter for
an end-contacted LL~\cite{bockrath},
\begin{equation}
 \alpha = \left( \bar{g}^{-1} - 1 \right) / 4
\end{equation}
where $\bar{g}$ is the effective Luttinger parameter for
crossed LL~\cite{komnik}
given by $\bar{g} = 2g$, we get $g = 0.23$ for CNT-1 and $g = 0.11$
for CNT-2.

For a LL, the temperature dependence of low-bias conductance is
also expected to show the power-law functional form,
$G$($V\approx0)\sim T^\alpha$. One way to exhibit this behavior is
to plot the scaled differential conductance, $T^{-\alpha} dI/dV$,
as a function of the scaled voltage, $eV/k_BT$~\cite{bockrath}. As
shown in the main panels of Fig. 2, the scaled differential
conductance curves for different temperatures fall well into a
single scaling curve, except for high bias voltage where the
differential conductance becomes saturated. Two CNTs showed
similar scaling behavior but with different exponent $\alpha$.
Such scaling behavior of the differential conductance curves is an
indication of probable LL behavior of the two CNTs in the sample.
The exponent $\alpha$ depends on the sample geometry.

The LL behavior of CNTs was further verified by the two-terminal
differential conductance curves of CNT in a crossed geometry
proposed by Komnik and Egger~\cite{komnik}. We have measured the
differential conductance of the CNT-1, $dI_{34}/dV_{34}$,
with applying bias voltage $V_{56}$ to the CNT-2. Fig. 3 (a) shows
the measured differential conductance curves with varying bias
voltage $V_{56}$ from -24 mV to +21.6 mV with the increment of 2.4
mV. For $V_{56}$ close to zero, typical differential conductance
curves with vanishing $dI/dV$ in a power of $V$ were shown. With
increasing the magnitude of $V_{56}$, the zero-bias conductance
increases rapidly and for $|V_{56}| > 12 \mbox{\rm mV}$, the
zero-bias differential conductance $dI_{34}/dV_{34}$ ($V_{34}$ =
0) switches from a dip to a peak. In addition the differential
conductance curve begins to exhibit two separated dips at finite
voltages. This characteristic feature agrees very well with the
theoretical prediction~\cite{komnik} and is one of experimental
evidences of the strong correlation between the two CNTs. Such a
strong dependence of the differential conductance
$dI_{34}/dV_{34}$ on the bias $V_{56}$ may not be easily
understood within the framework of the non-interacting electron
picture~\cite{buttiker}. The dip separation increases
monotonically with $|V_{56}|$. Following Komnik and
Egger~\cite{komnik}, the conductance dip should appear at
$|V_{56}| = |V_{34}|$ for two identical LLs with $g = 1/4$, a
special point where an exact $I$-$V$ curve was calculated. We have
plotted the dip position $V_{34}$ as a function of $V_{56}$ in
Fig. 3 (b). As expected, the peak position $V_{34}$ increases or
decreases linearly with $V_{56}$. Best fit gives the magnitude of
the linear slopes close to 0.21, about 5 times smaller than that
of the predicted ones for the two identical LLs. In our case, two
CNTs have different Luttinger parameters, $g$ = 0.23 for CNT-1 and
$g$ = 0.11 for CNT-2, respectively, which might be the origin of
the discrepancy.

We have also measured the current-voltage characteristics with
applying current to one CNT and measuring voltage drop on the
other CNT in the crossed geometry. Fig. 4 shows the
voltage-current characteristics, $V_{34}$-$I_{56}$, and the
off-diagonal differential resistance (ODR)-current curve,
$dV_{56}/dI_{34}$-$I_{34}$, measured at $T$ = 50 mK. A
noticeable feature is the highly-non-linear behavior of the ODR
together with the existence of the negative differential
resistance (NDR) at low bias region. This can be interpreted as
another indication of the existence of strong correlation between
two CNTs which are considered to be LLs. The interchange of the
voltage and the current leads gives almost identical
voltage-current characteristics. The NDR and asymmetry in the ODR
can be understood in a simple argument. The ODR can be written by
\begin{equation}
 R_{ab} = \frac{-G_{ab}}{ G_{aa}^2 - G_{ab}^2 } \;\;,
\end{equation}
where $G_{aa}$ and $G_{ab}$ are the diagonal and the off-diagonal
conductances given in Ref.\cite{komnik}. Here the indices
$a,b$ denote the current and the voltage probes in a given
measurement configuration, respectively. It can be shown that for
$g < 1/2$, $G_{ab}$ is non-zero and bound by $-G_0/2 < G_{ab} <
G_0/2$, where $G_0$ is the unit conductance of the system (in our
case $G_0=4e^2/h$), and can be asymmetric under the bias reversal.
Then $R_{ab}$ can be negative and also be asymmetric under the
bias reversal. These features are well shown in Fig. 4.

In summary, we have investigated electrical transport properties of
the crossed multi-wall CNTs. Each nanotube showed the tunneling density
of states vanishing with the power of bias and temperature at low energy
limit,
which is an evidence of the LL behavior. The differential conductance
curves of one CNT were disturbed significantly by applying bias voltage
to the other one in a crossed geometry. This characteristic feature is
an indication that strong correlation exists between the two crossed CNTs.
With increasing bias voltage, two dips appear in the differential
conductance curves and the dip separation increases linearly with the
bias voltage, which is consistent with the theoretical prediction for two crossed
LLs. Furthermore, the off-diagonal differential resistance exhibited
highly non-linear behavior with negative differential resistance. We
conclude that all these experimental results support the LL
behavior of multi-wall CNTs.

We thank G. Cuniberti, H.-W. Lee, and H. S. Sim 
for helpful discussions and comments. 
This work was supported by the MOST through Nano Structure
Project, Korea Research Foundation Grant (KRF-99-015-DP0128), BK21
project, and also by KRISS (Project No. 00-0502-100).

\begin{figure}
\caption{The scanning electron microscope image of the sample studied.
 For convenience, the electrical leads are numbered.}
\label{Fig. 1}
\end{figure}

\begin{figure}
\caption{The scaled conductance $T^{-\alpha}dI/dV$ as a function
of the scaled voltage $eV/k_BT$ with (a) $\alpha = 0.3$ for the
CNT horizontally placed in Fig. 1 (CNT-1) and (b) $\alpha=0.9$ for
the CNT vertically placed (CNT-2). Insets show the log-log plots
of the four-terminal differential conductance-voltage curves for
CNT-1 and CNT-2 at temperatures listed. The solid lines are guides
to show the power-law dependence of the tunneling density of
states. } \label{Fig. 2}
\end{figure}

\begin{figure}
\caption{(a) The two-terminal differential conductance-voltage
 ($dI_{34}/dV_{34}$ - $V_{34}$) curves of the CNT-1 with varying bias
voltage to the CNT-2, $V_{56}$, from -24 mV to +21.6 mV with the
increment of 2.4 mV at temperature $T$ = 50 mK. For clarity, each
curve was displaced vertically. (b) The dip position, $V_{34}$, in
the differential conductance-voltage curves of the CNT-1 as a
function of the bias voltage to the CNT-2, $V_{56}$. The solid
lines are linear fits to the data.} \label{Fig. 3}
\end{figure}

\begin{figure}
\caption{The off-diagonal voltage-current and differential
 resistance-current curves at temperature $T$ = 50 mK. We
 have applied bias current to the leads 3 and 4 and measured the
 voltage drops between the leads 5 and 6.}
\label{Fig. 4}
\end{figure}


\begin{references}
%
\bibitem{tarucha} S. Tarucha, T. Honda, and T. Saku, Solid State Comm.
 {\bf 94}, 413 (1995); F. P. Miliken, C. P. Umbach, and R. A. Webb,
 {\em ibid.} {\bf 97}, 309 (1995); A. M. Chang, L. N. Pfeiffer,
 and K. W. West, Phys. Rev. Lett. {\bf 77}, 2538 (1996);
 M. Grayson {\em et al.}, {\em ibid.} {\bf 80}, 1062 (1998);
 A. Yacoby {\em et al.}, {\em ibid.} {\bf 77}, 4612 (1996);
 O. M. Auslaender {\em et al.}, {\em ibid.} {\bf 84}, 1764 (2000);
 Q. Si, {\em ibid.} {\bf 81}, 3191 (1998); C. Winkelholz, R. Fazio,
 F. W. J. Hekking, and G. Sch\"on, {\em ibid.} {\bf 77}, 3200(1996).
\bibitem{bockrath} M. Bockrath {\em et al.}, Nature {\bf 397}, 598 (1999).
\bibitem{komnik} A. Komnik and R. Egger, Phys. Rev. Lett. {\bf 80}, 2881 (1998).
\bibitem{egger} R. Egger, Phys. Rev. Lett. {\bf 83}, 5547 (1999).
\bibitem{dekker} C. Dekker, Physics Today {\bf 52}, 22(1999);
  J. W. G. Wild\"oer {\em et al.}, Nature {\bf 391}, 59 (1998); 
  T. W. Odom, J.-L. Huang, P. Kim, and C. M. Lieber, {\em ibid.},
  62 (1998) ; L. C. Venema, {\em et al.}, Science {\bf 283}, 52 (1999);
 S. Frank, P. Poncharal, Z. L. Wang, W. A. de Heer, {\em ibid.} {\bf 280},
 1744 (1998); A. Bachtold, {\em et al.}, Nature {\bf 397}, 673 (1999);
 L. Langer {\em et al.}, Phys. Rev. Lett. {\bf 76}, 479 (1996);
 Z. Yao {\em et al.}, Nature {\bf 402}, 273 (1999);
 Z. Yao {\em et al.}. Phys. Rev. Lett. {\bf 84}, 2941 (2000);
 C. Sch\"onenberger, {\em et al.}, unpublished (cond-mat/9905114).
\bibitem{fuhrer} M. S. Fuhrer {\em et al.}, Science {\bf 288}, 494 (2000).
\bibitem{buttiker} M. B\"uttiker, Phys. Rev. Lett. {\bf 57}, 1761 (1986).
\bibitem{lee} J.-O. Lee {\em et al.}, to appear in Phys. Rev. B (2000).
%
\end{references}
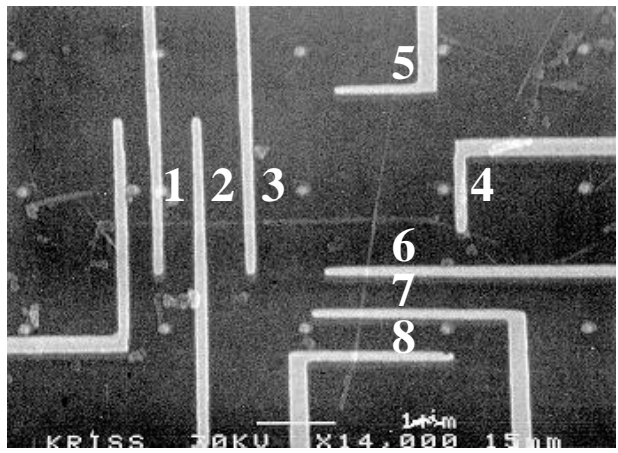
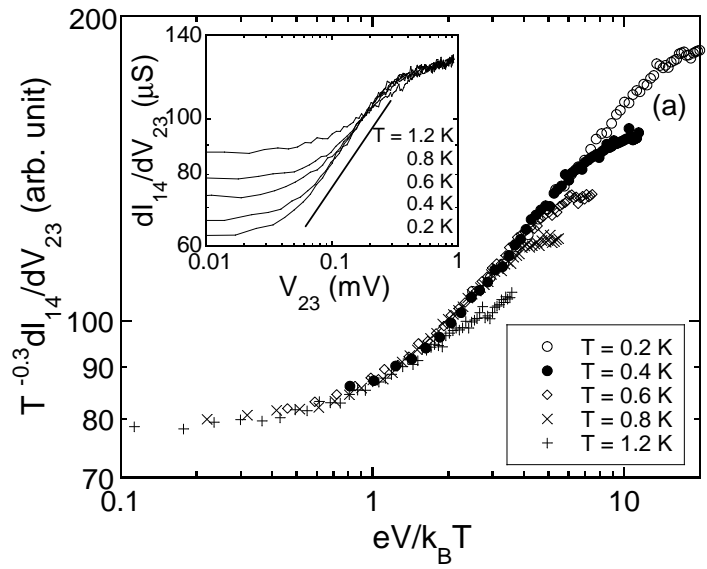
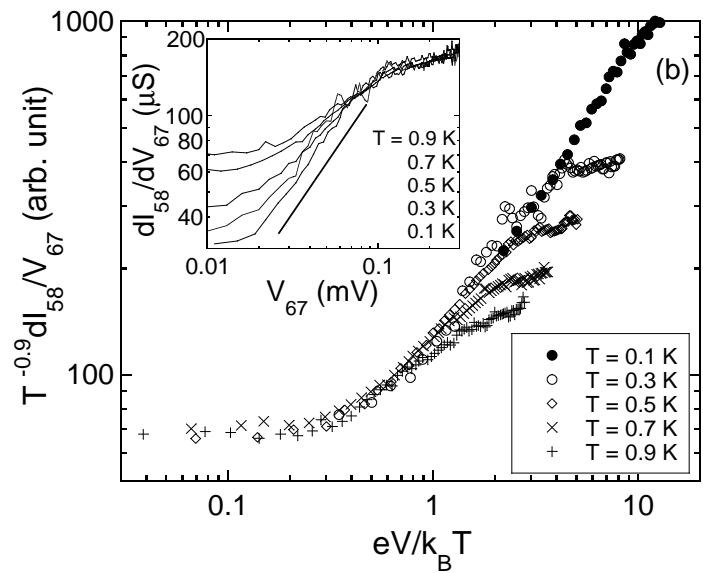
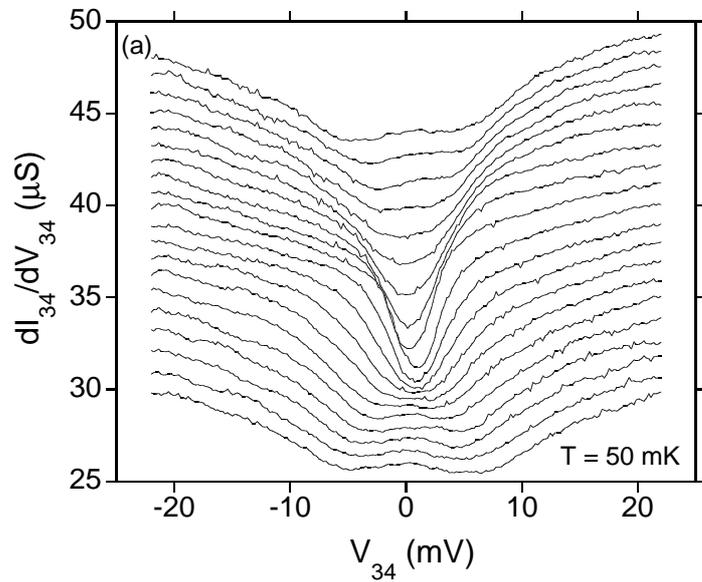
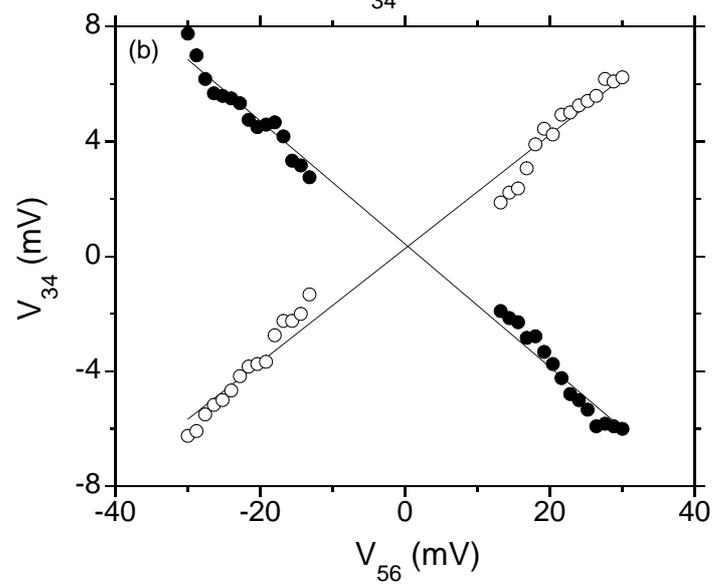
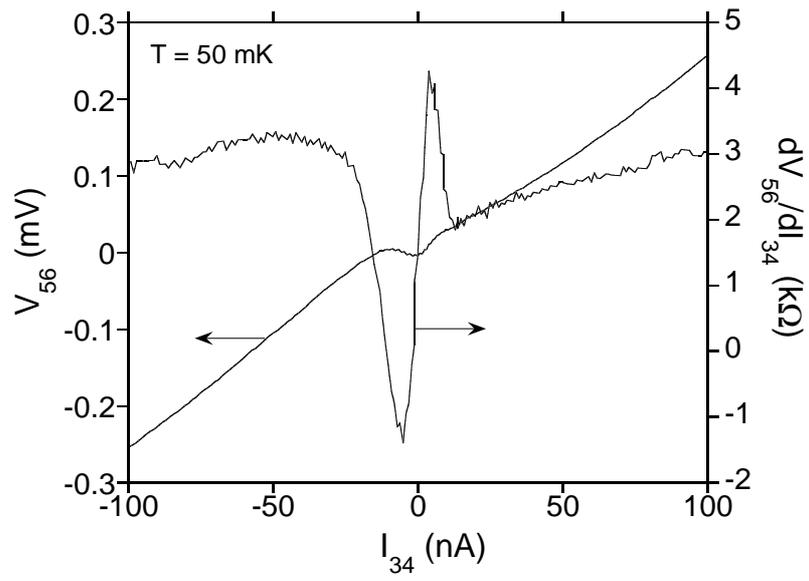
\end{document}